\documentclass[aip,jap,
reprint,
twocolumn,
amssymb,amsmath,fleqn]{revtex4-1}

\usepackage{bm}%
\usepackage[colorlinks=true,linkcolor=blue]{hyperref}%
\usepackage{graphicx}
\expandafter\ifx\csname package@font\endcsname\relax\else
 \expandafter\expandafter
 \expandafter\usepackage
 \expandafter\expandafter
 \expandafter{\csname package@font\endcsname}%
\fi
\usepackage{color}

\newcommand{\blue}[1]{{\color{blue} #1 }}

\begin{document}

\title{First-principles study on the magnetic properties of ordered Nd$_{6}$(Fe,Ga)$_{14}$ alloys }

\author{Kazushige~\surname{Hyodo}}
\affiliation{Department of Applied Physics, Tohoku University, Aoba 6-6-05, Aoba-ku, Sendai, Japan}

\author{Yuta~\surname{Toga}}
\affiliation{ESICMM, National Institute for Materials Science, Sengen 1-2-1, Tsukuba, Japan}

\author{Akimasa~\surname{Sakuma}}
\affiliation{Department of Applied Physics, Tohoku University, Aoba 6-6-05, Aoba-ku, Sendai, Japan}
\email{E-mail address: sakuma@solid.apph.tohoku.ac.jp}
\begin{abstract}
We studied the stable magnetic structure of ordered Nd$_{6}$Fe$_{14-x}$Ga$_x$ ($x = 0, 1)$ alloys, which appears in the grain-boundary (GB) phase of Nd-Fe-B permanent magnets, using first-principles techniques.
Slight Ga doping ($x = 1$) was shown to contribute to the stabilization of an anti-ferromagnetic (AF) state, whereas the non-doped case ($x = 0$) was revealed to favor ferromagnetic state rather than AF state with a slight energy difference. 
\end{abstract}

\maketitle

The intermetallic compounds $R_6$Fe$_{14-x}M_x$ ($R$=rare earth, $M$=Si, Ga, Al, Ge, Cu etc.) have attracted attention due to their interesting properties such as meta-magnetic transition at a few Tesla of magnetic field\cite{Jonen1997,Jonen1999} and large magnetic anisotropy field larger than 7 T.\cite{Li1990}
Also, in technological viewpoints, these alloys were intensively ivestigated because their existence as grain boundary (GB) phase in Nd-Fe-B permanent magnets enhances the coercive force $H_{\mathrm{c}}$,\cite{Schrey,Kajitani} and they absorb large amount of hydrogens without any change of symmetry.\cite{Coey,Pourarian} Quite recently, the effects of Nd$_6$Fe$_{14-x}$Ga$_x$ as a GB phase in Nd-Fe-B magnets has been revisited,\cite{Sasaki} since the higher $H_{\mathrm{c}}$ of Nd-Fe-B magnets is in great demand for realizing more energy-efficient motors. In particular, motors in recent electric vehicles that operate under high temperature require a larger $H_{\mathrm{c}}$ to suppress thermal fluctuations of the magnetization.  

Besides these attractive properties, the magnetic structure has not yet been established and has so far been controversy.
Experimentally, various measurements were performed, such as M$\mathrm{\ddot{o}}$ssbauer measurement,\cite{Kajitani,Knoch,Coey,Jonen1997,Hu,Weitzer} neutron\cite{Yan,Schobinger} and X-ray diffractions.\cite{Allemand,Hu,Knoch,Groot,Weitzer}
The neutron and some measurements\cite{Schobinger,Allemand,Kajitani,Groot} proposed that the $R_6$Fe$_{14-x}M_x$ form antiferromagnetic (AF) structure, while other experiments suggested ferri-\cite{Hu,Knoch,Yan} or ferromagnetic structures.\cite{Li1990}
In addition, the reported magnetic moments on Fe sites were different depending on the experimental methods.\cite{Schobinger99,Isnard}
In order to gain insight into the magnetism of this alloy, the first principles calculations for the electronic and magnetic structures could be helpful. 

In the present study, we focus on the Nd$_6$Fe$_{13}$Ga alloy and investigated the magnetic properties of this system using the first-principles technique. The main purpose of this work is to determine the magnetic structure including magnetic moments on each Fe site and the next is to examine its stability, in order to provide helpful information to understand the role of these alloys as a GB phase in Nd-Fe-B magnets. For the latter aim, we are concerned with the effects of Ga atoms on the magnetism of this system. To see this, we examined also the magnetic structure of hypothetical Nd$_6$Fe$_{14}$ alloy which may be unstable to exist alone, and compared it with that of Nd$_6$Fe$_{13}$Ga.

\begin{figure}
\includegraphics[width=7.5cm]{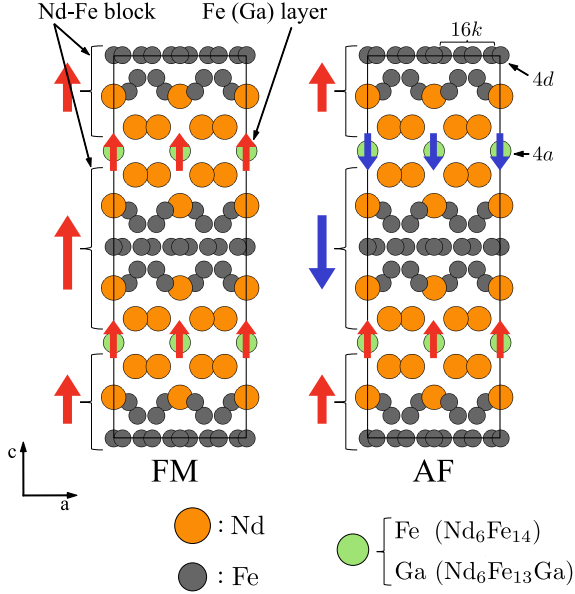}
\caption{Unit cell and candidates of stable magnetic structures of ordered Nd$_{6}$Fe$_{14}$ and Nd$_{6}$Fe$_{13}$Ga. The green circles indicate the atomic positions denoted by 4$a$, which are occupied by Ga or Fe atoms in Nd$_{6}$Fe$_{13}$Ga or Nd$_{6}$Fe$_{14}$, respectively. The top, botom and central layers are constituted of 4$d$ and 16$k$ Fe atoms.
%The employed atomic positions are the observed value of Nd$_{6}$Fe$_{13}$Ga in the previous experiment. \cite{19}. 
These crystal structures were drawn using VESTA.\cite{VESTA}}%in Nd$_{6}$Fe$_{13}$Si alloys.
\label{f2}
\end{figure}

Since the crystal structure of Nd$_6$Fe$_{13}$Ga is quite complex and the number of atoms in the unit cell is so large, there exist infinite possibilities in magnetic structure. Therefore, we concentrate ourselves to the AF structure proposed by neutron measurement\cite{Schobinger} as a candidate of the magnetic structure, and investigate the stability of the AF structure.
It was found, as a result of the present studies, that Nd$_6$Fe$_{14}$ alloy favors ferromagnetic state rather than AF state with a slight energy difference, and the substitution of Fe atoms by Ga atoms makes the AF state much stable, leading to Nd$_6$Fe$_{13}$Ga.  
We also revealed that this stable AF state originated from the anti-parallel magnetic coupling between the neighboring Nd-Fe blocks shown in Fig.~\ref{f2} and few magnetism of the doped Ga contributed to the stabilization of this magnetic state.

The unit cell of the ordered Nd$_{6}$Fe$_{14-x}$Ga$_{x}$ ($x=0,1$) system consists of 80 atoms, as shown in Fig.~\ref{f2}, according to a previous experimental report.\cite{Li}
When $x = 1$, the doped Ga atoms replace the 4$a$ Fe atoms (see Fig.~\ref{f2}).
We adopted the displayed FM and AF states shown in Fig.~\ref{f2} as candidates for stable magnetic states. 
The characteristic magnetic structure of the AF state is based on the observed AF state in Nd$_6$Fe$_{12}$Ga$_{2}$ alloys.\cite{Schobinger} %using neutron diffractions
In both states, the magnetic moments of the internal atoms in the Nd-Fe blocks and Fe(Ga) layers are parallel, as shown by the arrow in Fig.~\ref{f2}. 
In contrast, the magnetic moments of the adjacent Nd-Fe block are parallel in the FM state and anti-parallel in the AF state. 
%The stable magnetic state is determined by a direct comparison of the total energy between the FM and AF states. 

We used the Vienna ab initio simulation package (VASP 5.4.1) as a first-principles calculation method for obtaining the total energy.\cite{VASP} 
The cell volume, lattice constants, and internal atom positions were redefined from the experimental values \cite{Li} using the self-consistent relaxation operation in each composition and magnetic state. 
The cut-off energy is $334.9$ eV, and the Monkhorst--Pack $\bm{k}$-point meshes are $5\times 5 \times 3$ in collinear calculations and $3\times 3\times1$ in non-collinear calculations. 
A self-consistent electronic structure was obtained for valence electrons, except for the $4f$ electrons in the Nd atoms, which were treated as core electrons in this study. 
The ionic potentials is described by the plane-augmented-wave (PAW) method,\cite{PAW1,PAW2} and 
the exchange-correlation energy of the valence electrons is represented within the generalized gradient approximation (GGA), whose specific form was given by Ceperly and Alder and parametrized by Perdew et al.\cite{GGA}

%Table \ref{t2} (a) shows the calculated energy of the FM and AF states in the unit cell of Nd$_{6}$Fe$_{14-x}$Ga$_{x}$ $(x=0, 1)$. 

Table \ref{t111} (a) shows the lattice constant of each magnetic state after the relaxation process in both alloys. 
%The result of relaxed lattice constant of each magnetic states is also shown in \ref{t2} (b) about the both compositions. 
The obtained lattice constants are almost equal between the two alloys and the two magnetic states.  
%We confirmed that the obtained relaxed lattice constant is almost equal between the FM and AF states in both systems. 
%さらにNd6Fe14,Nd6Fe13Ga間やFM,AF間で比較した結果を入れる(FMだとわずかにc軸長が伸びる)
In addition, the obtained relaxed value of the AF state in Nd$_{6}$Fe$_{13}$Ga is almost the same as the experimental value ($a$ = 8.069~\AA \hspace{1ex} and $c$ = 22.937~\AA,\cite{Li} the differences are about $0.8\%$).

%$E_{\mathrm{FM}}$ ($E_{\mathrm{AF}}$) indicates the calculated energy of the FM (AF) state, 

%
%estimated exchange constant $J$, which is derived from Eq. (\ref{eq:2}). %, between the different magnetic domains of the AF state. 
%The value of $S$ in Eq. (\ref{eq:2}) is the relaxed value of the AF state (the difference of the square area between the FM and AF states is less than 1\% in both alloys, as shown in table \ref{t2}). 

In table \ref{t111} (b), using the relaxed atomic positions, we show the calculated energy of the FM and AF states ($E_{\mathrm{FM}}$, $E_{\mathrm{AF}}$) and the difference of them, $\Delta E = E_{\mathrm{FM}} - E_{\mathrm{AF}}$.
Additionally in this study, for convenience in comparison among different systems, we defined the interface magnetic coupling energy:
\begin{align}
J = \frac{E_{\mathrm{FM}} - E_{\mathrm{AF}}}{2S}, \label{eq:2}
\end{align}
where $S$ is the interface area of the Nd-Fe block in the case of AF state (the difference of the square area between the FM and AF states is less than 1\% in both alloys, as shown in table~\ref{t111} (a)).
Note that two interfaces between Nd-Fe blocks exist in the unit cell.
Positive (negative) $J$ indicates preference of the AF (FM) state over the FM (AF) state. 
These results in table \ref{t111} (b) indicated that the Nd$_6$Fe$_{13}$Ga prefer AF states whereas the Nd$_6$Fe$_{14}$ favours FM state with a slight energy difference.

%From Eq. (\ref{eq:1}), $J$ is derived from the first-principles calculation results: %obtained energies; 
%\begin{align}
%J = \frac{E_{\mathrm{FM}} - E_{\mathrm{AF}}}{2S}, \label{eq:2}
%\end{align}
%where $E_{\mathrm{FM}}$ ($E_{\mathrm{AF}}$) indicates the calculated energy of the FM (AF) state, and $S$ is the interface area of the Nd-Fe block in the case of AF state (the difference of the square area between the FM and AF states is less than 1\% in both alloys, as shown in table~\ref{t111} (a)).
%Note that two interfaces between Nd-Fe blocks exist in the unit cell.
%%
%Positive (negative) $J$ indicates preference of the AF (FM) state over the FM (AF) state. 

%#############################################
%#############################################
\begin{table}
\caption{(a): Obtained lattice constants along the $a$- and $c$-axes after the relaxation process in the AF and FM states of Nd$_{6}$Fe$_{13}$Ga and Nd$_{6}$Fe$_{14}$. The lattice constant along the $b$-axis is same as that along the $a$-axis due to the symmetry of this system.
(b): Calculated energies in the FM ($E_{\mathrm{FM}}$) and the AF  ($E_{\mathrm{AF}}$) states for the unit cell shown in Fig.~\ref{f2} and the difference between them, $\Delta E$.
Interface coupling energy, $J$, defined by Eq. (\ref{eq:2}).
}
\label{t111}
%#############################################
\begin{center}
(a)
\end{center}
\vspace{-0.5em}
\begin{tabular}{ccc}
\hline
 & $a$ (\AA) & $c$ (\AA) \\  
\hline
Nd$_{6}$Fe$_{13}$Ga(FM)  & $8.004$ & $23.209$ \\
Nd$_{6}$Fe$_{13}$Ga(AF) & $8.006$ & $23.119$ \\
Nd$_{6}$Fe$_{14}$(FM) & $7.986$ & $22.959$ \\
Nd$_{6}$Fe$_{14}$(AF) & $8.004$ & $22.773$ \\
%Nd$_{6}$Fe$_{13}$Ga(FM)  & $8.0038$ & $23.2090$ \\
%Nd$_{6}$Fe$_{13}$Ga(AF) & $8.0056$ & $23.1190$ \\
%Nd$_{6}$Fe$_{14}$(FM) & $7.9846$ & $22.9593$ \\
%Nd$_{6}$Fe$_{14}$(AF) & $8.0035$ & $22.7726$ \\
\hline
\end{tabular}
%#############################################
\begin{center}
(b)
\end{center}
\vspace{-0.5em}
%the difference of energy between the two states  ($\Delta E$) in the both alloys. The exchange constant $J$ between the Nd-Fe blocks derived in Eq. () is also shown}
\begin{tabular}{crrrr}
\hline
& $E_{\mathrm{FM}}$ (eV) & $E_{\mathrm{AF}}$ (eV) & $\Delta E$ (eV)  &$J$ (mJ/m$^2$) \\  
\hline
Nd$_{6}$Fe$_{13}$Ga & $-556.752$ & $-556.891$ &  $0.139$ & $8.7 $ \\
Nd$_{6}$Fe$_{14}$   & $-568.867$ & $-568.846$ & $-0.021$ & $-1.3$ \\
%Nd$_{6}$Fe$_{13}$Ga  & -556.7522 & -556.8919 & 0.139 & 0.0347 \\
%Nd$_{6}$Fe$_{14}$  & -568.867 & -568.846 & -0.021 & -0.0053 \\
\hline
\end{tabular}

\end{table}
%#############################################
%#############################################

%#############################################
\begin{figure}
\qquad(a)
\begin{center}
\includegraphics[height=5.2cm]{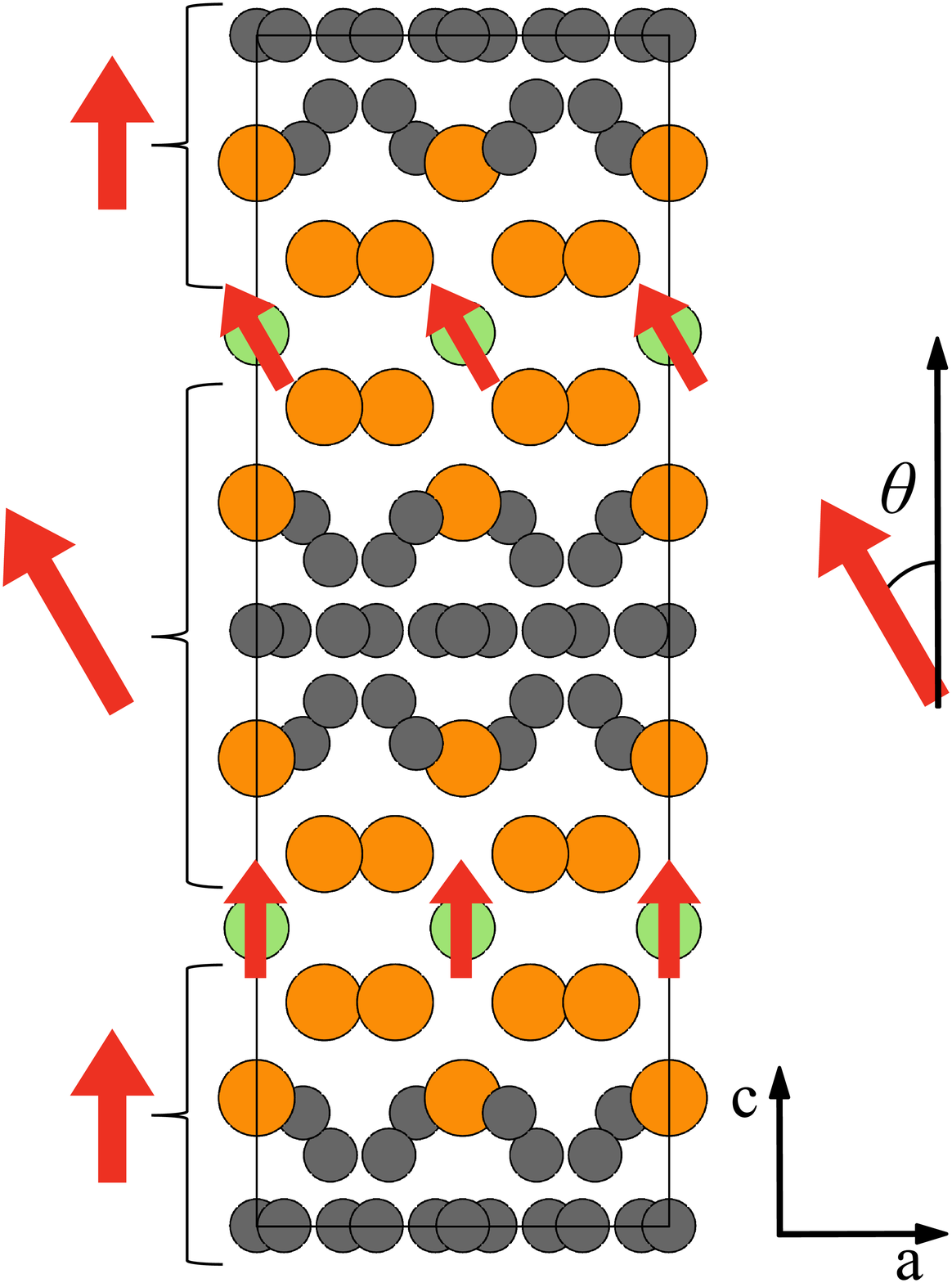}
\end{center}
\qquad(b)
\begin{center}
\includegraphics[width=8cm]{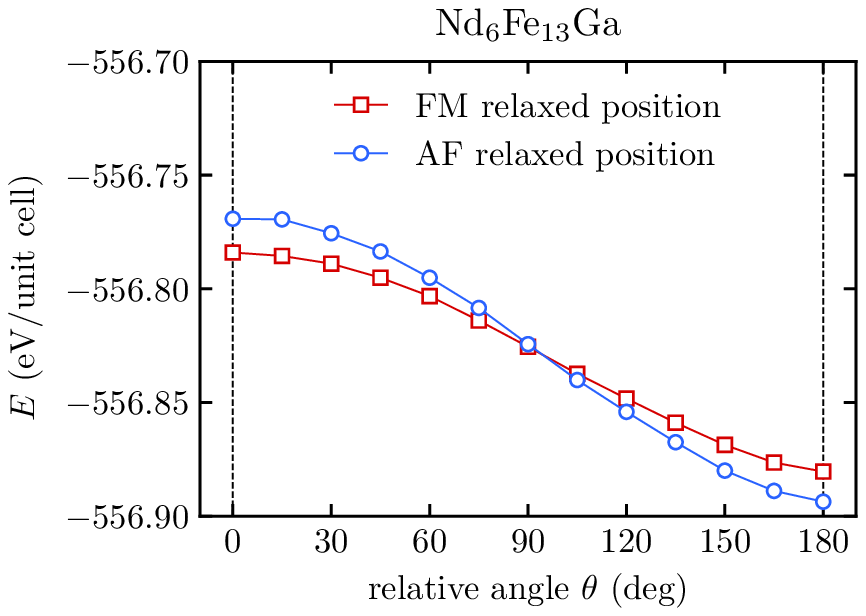}
\end{center}
\qquad(c)
\begin{center}
\includegraphics[width=8cm]{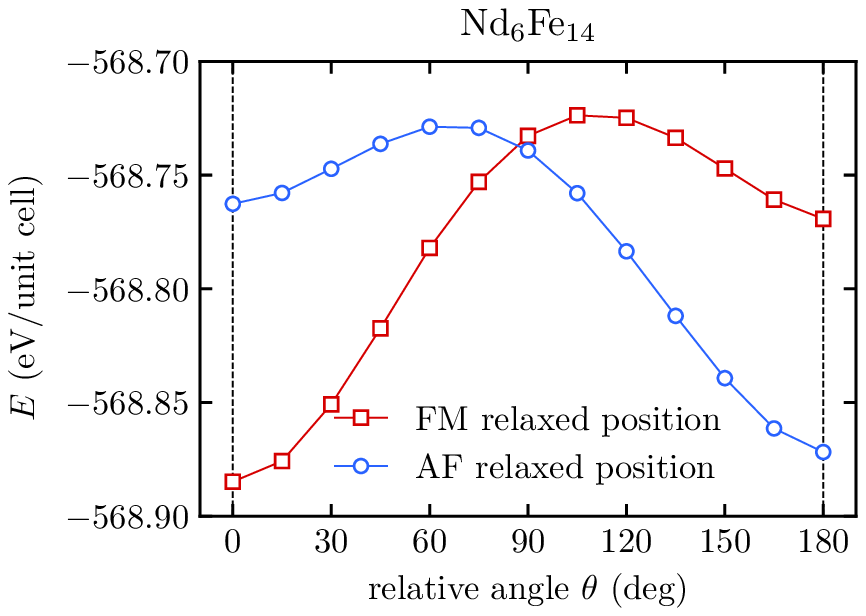}
\end{center}
\caption{(a) Our considered magnetic configuration, where the magnetic moments in half of the Nd-Fe blocks and Fe(Ga) layers have a relative angle $\theta$ to the magnetic moments of the other blocks and other layers. (b) Calculated energy of the unit cell in Nd$_{6}$Fe$_{13}$Ga in each condition of $\theta$. (c):  Calculated energy of the unit cell in Nd$_{6}$Fe$_{14}$ in each condition of $\theta$. }
\label{f6}
\end{figure}

To further understand the energetic preference between the AF and FM states in the Nd$_{6}$Fe$_{14-x}$Ga$_{x}$ ($x = 0, 1$) system, we calculated the energy of assumed non-collinear magnetic moments, which continuously changed from the FM to AF state. 
Figure \ref{f6} (a) shows our considered magnetic structure, where the magnetic moments in half of the Fe-Nd blocks and Fe(Ga)-layers are rotated by an angle $\theta$ from those in the other blocks and layers. 
Figure \ref{f6} (b) and (c) present the obtained energy of Nd$_{6}$Fe$_{13}$Ga and Nd$_{6}$Fe$_{14}$ as a function of $\theta$. 
The red and blue lines shows the calculation results under the relaxed atomic positions in the FM and AF states, respectively. 
The case of $\theta = 0^{\circ}$ ($\theta = 180^{\circ}$) in each figure corresponds to the FM (AF) state in Fig.~\ref{f2}. 
Note that the energies in these cases are a slight decrease from the results in Table~\ref{t111} (b), which is owing to the non-collinear spin configurations.
Even if total moment is along $c$-axis, the spins are not completely collinear but slightly inclined to each other, and reduce the energies.
Although the non-collinear calculation condition may reduce the results of $|J|$ by about 30\% at maximum, the change does not affect the discussion in this study.

In Nd$_{6}$Fe$_{13}$Ga (Fig.~\ref{f6}~(b)), the $\theta = 180^{\circ}$ and $\theta = 0^{\circ}$ cases correspond to the minimum and maximum points, respectively, and the obtained energy monotonically decreases with the increase of $\theta$. 
In addition, this trend is shown in both lines, which indicate that the magnetic structure shifts to the AF state, regardless of the atomic positions. 
Thus, we confirmed that the AF state was preferred over the FM state in Nd$_{6}$Fe$_{13}$Ga. 
In contrast, in Nd$_{6}$Fe$_{14}$ (Fig.~\ref{f6} (c)), the dependence of energy on $\theta$ was found to differ between cases with different atomic positions; if a stable atomic position is adopted in the FM state, the FM state ($\theta = 0^{\circ}$) becomes the ground state, whereas the AF state ($\theta = 180^{\circ}$) becomes the ground state for atomic positions in the AF state. 
%This result indicates that the stable magnetic structure is easily changed depending on the atomic position. Whereas the magnetic structure is less stable compared to Nd$_{6}$Fe$_{13}$Ga as seen above, energies in the two lines show that the FM state is rather stable in Nd$_{6}$Fe$_{14}$.  
This result indicates that the stable magnetic structure easily changes depending on the atomic position in Nd$_{6}$Fe$_{14}$, even though the FM state is rather stable compared to the AF state in this structure, as shown in table~\ref{t111} (b).

\begin{table}
\caption{Calculated energy of the FM and AF states ($E_{\mathrm{FM}}$ and $E_{\mathrm{AF}}$) for the unit cell, the difference between them $\Delta E = E_{\mathrm{FM}} - E_{\mathrm{AF}}$, and the interface coupling energy $J$ in Nd$_{6}$Fe$_{13}M$ ($M$ = Si, Al, empty).}%in Nd$_{6}$Fe$_{13}$M (M = Ga, Si, Al, empty). The energy difference $\Delta E$ between the two states and the exchange constant $J$ evaluated from Eq. (\ref{eq:2}) is also shown. }
\label{t3}
\begin{tabular}{crrrr}
\hline
$M$ & $E_{\mathrm{FM}}$(eV) & $E_{\mathrm{AF}}$(eV) & $\Delta E$ (eV) & $J$ (mJ/m$^2$) \\
\hline
% \red{\sout{Ga}} & \red{\sout{  $ -556.752 $ }} & \red{\sout{ $ -556.891 $ }}& \red{\sout{ $0.139$ }}& \red{\sout{ $0.035$ }}\\
 Al & $ -557.908 $ & $ -558.056 $ & $0.148$ & $  9.2$ \\
 Si & $ -567.891 $ & $ -568.149 $ & $0.258$ & $ 16.2$ \\
 empty & $-536.742$ & $-536.875$  & $0.133$ & $  8.3$ \\
%$M$ = empty & $-536.7422$ & $-536.875$ & $0.033$ \\
%M = Ga & $-556.752$ & $-556.891$ & $0.139$ & $0.035$  \\
%M = Al & $ -557.908 $ & $ -558.056 $ & $0.148$ & $0.037$  \\
%M = Si & $-567.891 $ & $ -568.149 $ & $$0.258$$ & $0.064$  \\
%M=empty & $-536.742$ & $-536.875$ & $0.133$ & $0.033$ \\
\hline
\end{tabular}
\end{table}

Next, we investigated the role of Ga atoms on the AF structure in Nd$_{6}$Fe$_{13}$Ga.
We focused on the almost non-magnetism of the doped Ga atom of Nd$_{6}$Fe$_{13}$Ga, whose magnetic moments were $0.0\mu_{\mathrm{B}}\rm/atom$ (see later in Table~\ref{tbl:moment}) in both the AF and FM states, respectively. 
From this result, it would be reasonable to presume that the stable AF state in Nd$_{6}$Fe$_{13}$Ga (Fig.~\ref{f2}) is almost dominated by the AF coupling between the neighboring Nd-Fe blocks and hardly relates to the magnetic interaction mediated by Ga atoms. 
To demonstrate this hypothesis, we compared the stability of the AF state in Nd$_{6}$Fe$_{13}M$, where the Ga atoms in Nd$_{6}$Fe$_{13}$Ga were replaced by some non-magnetic $M$ atoms or empty space. 
Table~\ref{t3} shows the calculated energies of the FM and AF states and the evaluated interface coupling $J$ using Eq. (\ref{eq:2}) in Nd$_{6}$Fe$_{13}M$ ($M$=Al, Si, empty).
%The result of M=Ga is same as that of table 1, and the result of M = Al, Si is obtained from the individual structure after the relaxation process, which was started from the experimental value of Nd$_{6}$Fe$_{13}$Si\cite{19}. 
The results of $M$ = Al, Si were calculated from the relaxed atomic positions, which was obtained with respect to each system and magnetic state. %, where the relaxation process is performed individually. 
In the case of $M$ = empty, we used the relaxed atomic position of Nd$_{6}$Fe$_{13}$Ga and removed the Ga atoms with holding the other atomic positions fixed. 
%\red{\sout{The $S$ value for evaluating $J$ in Eq. (\ref{eq:2}) was used the value of the AF state as well as the calculation of $J$ in table \ref{t2} (b).}}

We found that all obtained $J$ values were positive values. 
The preference of the AF state in the case of %M = Al, Si 
$M$ = Si, as well as the case of Nd$_{6}$Fe$_{13}$Ga, coincides with the observed result in previous experiments.\cite{Groot,Allemand}
In addition, the obtained $J$ values are similar between these systems.
Especially, the case of $M$ = empty has almost same value as Nd$_{6}$Fe$_{13}$Ga under the same structure without Ga.
From these results, we conclude that the role of Ga is only as a spacer and the stability of the AF state in Nd$_{6}$Fe$_{13}$Ga mainly originates from the anti-parallel coupling between the neighboring Nd-Fe blocks. 
At this stage, we consider that this anti-parallel coupling is mainly attributed to a kind of kinetic exchnage interaction (such as RKKY interaction in most RE metals) between the Nd layers on both sides of the Ga 4$a$ layer, which is mediated by $s$-  or $p$- electrons. 

%through the non-magnetic spacer of Ga atoms. 

%#############################################
\begin{table}
\caption{
Theoretical (this work, used Bader charge analysis\cite{bader_impl,bader}) and previous experimental (M\"ossbauer measurement (MS) or neutron diffractions (ND)) results of the amplitude of magnetic moment, $m_s$, for each Fe ion in some $R_6$Fe$_{13}M$ systems.
All theoretical results were evaluated in the stable (only Nd$_{6}$Fe$_{14}$ is the FM, and the others are the AF) states and in the relaxed atomic positions.
}
\label{tbl:moment}
\begin{tabular}{c|cccc|c}
\hline 
\multicolumn{1}{c}{$R_6$Fe$_{13}M$} & \multicolumn{5}{c}{$m_s$ ($\mu_B$)} \\ \hline \hline
  Theory  & \multicolumn{4}{c|}{Fe} &  $M$  \\  \cline{2-6}
  & 4$d$ & 16$k$ & 16$l_1$ & 16$l_2$ & 4$a$    \\  
\hline  
Nd$_{6}$Fe$_{13}$Ga & 1.95 & 2.24 & 2.15 & 2.28 & 0.00(3)   \\
Nd$_{6}$Fe$_{13}$Al & 1.96 & 2.25 & 2.16 & 2.30 & 0.00(4)  \\
Nd$_{6}$Fe$_{13}$Si & 1.97 & 2.23 & 2.15 & 2.30 & 0.00(3)  \\
Pr$_{6}$Fe$_{13}$Si & 2.05 & 2.27 & 2.17 & 2.35 & 0.00(2)  \\
Nd$_{6}$Fe$_{14}$   & 1.94 & 2.23 & 2.15 & 2.25 & 2.31     \\ % lu(4a)=1.3005
\hline\hline 
\multicolumn{1}{c}{Experiment} \\ \hline
Nd$_{6}$Fe$_{13}$Si@2K (ND)\cite{Isnard} & 2.8 & 2.6 & 2.4 & 1.8 & -    \\
Nd$_{6}$Fe$_{13}$Si@4.2K (MS)\cite{Isnard} & 2.5 & 2.3 & 2.1 & 1.6 & -  \\
\hline
Pr$_6$Fe$_{13}$Si@1.5K (ND)\cite{Schobinger99} & 0.9 & 2.0 & 2.1 & 2.1 & - \\
Nd$_6$Fe$_{13}$Au@1.5K (ND)\cite{Schobinger99} & 1.3 & 2.6 & 2.6 & 2.6 & - \\
\hline 

\end{tabular}
\end{table}
%#############################################

\begin{figure}
\includegraphics[width=8cm]{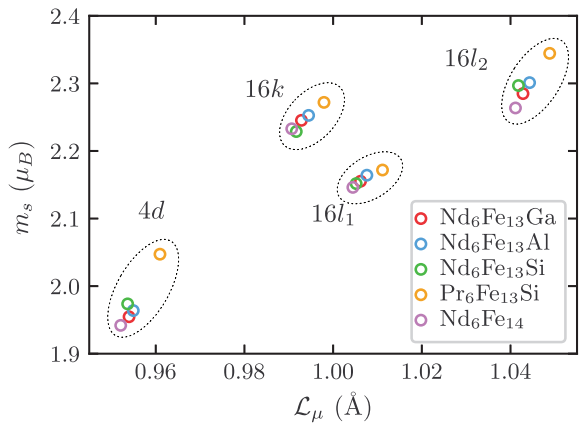}
\caption{The relationship between the canonical bond length, $\mathcal{L}_\mu$, definded by Eq.~\eqref{eq:canonical} and the magnetic moment, $m_s$, for each of the atoms and the substances.}
\label{fig:length}
\end{figure}

For more detailed studies of the substitution effect of the Ga $4a$ atoms,
the calculated magnetic moments, $m_s$, on each site for $M$=Ga, Al, Si, and Fe are listed in Table~\ref{tbl:moment},
together with the experimental data, for comparison.
And, we also added the calculation result of a Pr$_{6}$Fe$_{13}$Si system to compared with the experiment for the same substance.
One may notice that the Fe magnetic moments at each site have almost same values irrespective of substance except for the Pr$_{6}$Fe$_{13}$Si.
The slightly differences of the Pr$_{6}$Fe$_{13}$Si is due to changes in the space of each Fe site (see later and Fig.~\ref{fig:length}).
On the other hand, the remarkable difference can be found in the moment on 4$d$ Fe site between the calculated and experimental values; the moment on the 4$d$ site is the smallest in the calculated and the two experimental data at the bottom of Table~\ref{tbl:moment},
while one is the largest value in the other two experimental data.
We have confirmed that the trend in which the moment on 4$d$ site is the smallest can be seen also in other calculation method (Korringa-Kohn-Rostoker method) for electronic structure.\cite{doi_kkr}

%##################################
%##################################

Within the framework of electronic structure calculations, we take the trend owing to the situation that the 4$d$ Fe moments are surrounded relatively densely by other Fe sites such as 16$k$, 16$l_1$ and 16$l_2$ sites, resulting in strong itinerant character and the small magnetic moments.
To analyze this, we adopted a canonical bond length proposed by Harashima et al.\cite{harashima} as follow:
%##################################
\begin{eqnarray}
\mathcal{L}_\mu=\frac{1}{2}\left( \sum_\nu |\bm{r}_\mu-\bm{r}_\nu|^{-10} \right)^{-\frac{1}{10}},
\label{eq:canonical}
\end{eqnarray}
%##################################
here $\mu$ and $\nu$ are atom indices, and $\bm{r}$ is an atom position vector.
This length is based on the canonical band theory\cite{canonical} and formulated from $d$-$d$ hopping integral.
Since magnetic moment is mainly contributed by the $d$-electrons, thus $\mathcal{L}_\mu$ is reasonable to compare the atomic spaces with the magnetic moments.

Figure \ref{fig:length} shows clearly positive correlation between $\mathcal{L}_\mu$ and $m_s$ for each of the atoms and the substances.
For Pr$_6$Fe$_{13}$Si, it can be undertood that the increase of $\mathcal{L}_\mu$ reflect the slightly difference of $m_s$ in Table~\ref{tbl:moment}.
The canonical length is not possible to explain the reverse magnitude relationship between 16$k$ and 16$l_{1}$,
%to clarify this need to consider the effect of detailed electronic structure around the atoms.
%
%However,
whereas $\mathcal{L}_{4d}$ is definitely smaller than the others, which support that the moment on 4$d$ site is the smallest in Table~\ref{tbl:moment}.

%2E(nd-nd)+2E(nd-fe)=0.12
%4         4        =0.24
%          4E(nd-fe)=0.4
%            
%4(nd-nd)=-0.16    4(nd-fe)= 0.4
	
%Moreover, similar values are obtained in the $M$ = empty case, where the energy difference is considered to originate from only magnetic couplings between the Nd-Fe blocks. 
%From this result, we conclude that the stability of the AF state in Nd$_{6}$Fe$_{13}$Ga mainly originates from the anti-parallel coupling between the neighboring Nd-Fe blocks through the non-magnetic spacer of Ga atoms. 

%\{\sout{In addition, from the $J$ value in Nd$_{6}$Fe$_{13}$, we can estimate the exchange coupling between the neighboring Nd-Fe blocks as $J_{\mathrm{NdFe/NdFe}} \sim 0.033$ J/m$^2$.}}

%##################################
\begin{figure}
\begin{center}
\qquad(a)
\end{center}
\includegraphics[height=5.2cm]{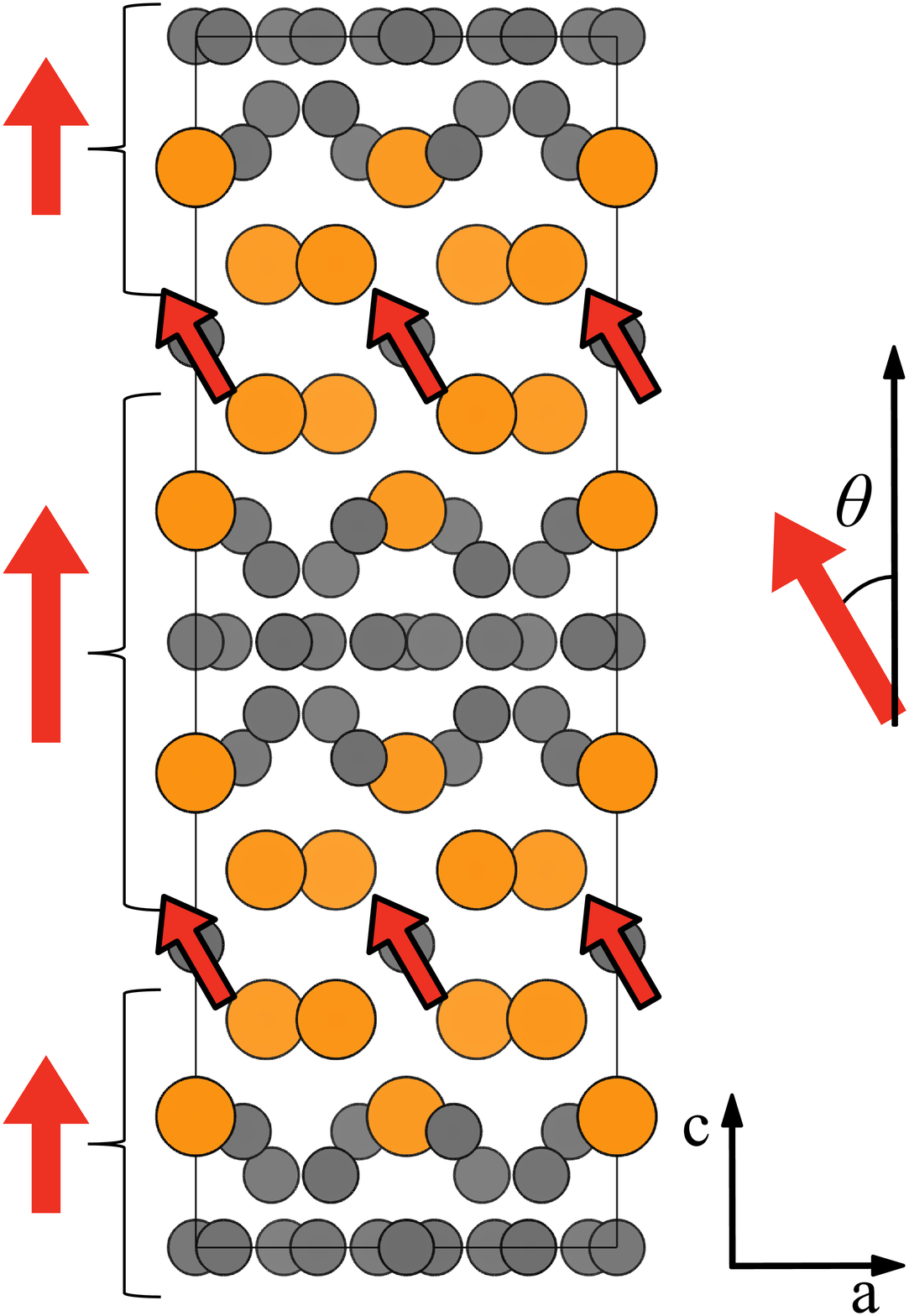}
\begin{center}
\qquad(b)
\end{center}
\includegraphics[width=8cm]{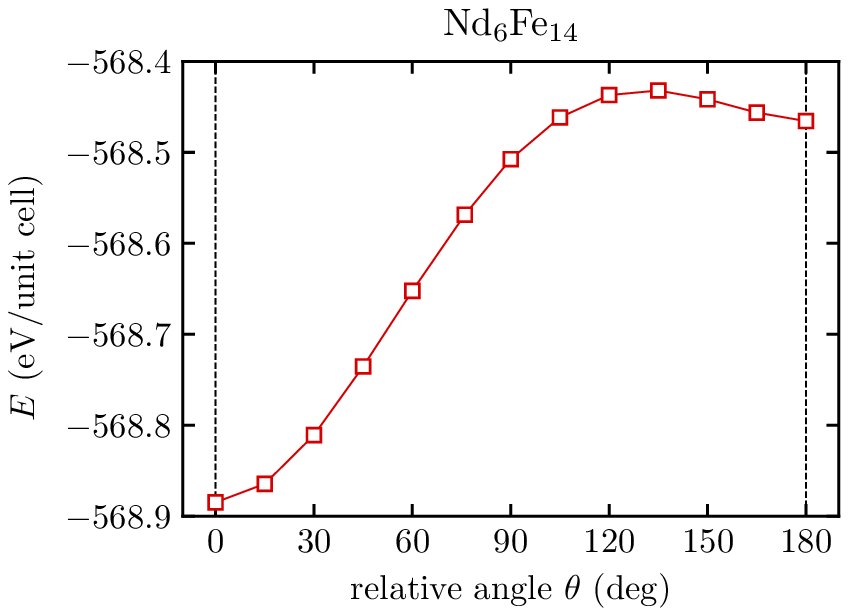}
\caption{(a): Assumed magnetic structure in Nd$_{6}$Fe$_{14}$, where the magnetic moments in the Fe layer have a relative angle $\theta$ to those in the Nd-Fe blocks. (b): Calculated energy per unit cell in Nd$_{6}$Fe$_{14}$ as a function of $\theta$. }
\label{f8}
\end{figure}
%##################################

Finally, we examined the origin of competition between the FM and AF states in the Nd$_{6}$Fe$_{14}$ system. 
From Table~\ref{tbl:moment} and Fig.~\ref{fig:length}, it is confirmed that the Nd$_{6}$Fe$_{14}$ is almost same as the Nd$_{6}$Fe$_{13}$Ga regarding $m_s$ and $\mathcal{L}_\mu$ except for $m_s$ of 4$a$ site.
The difference of magnetism between them mainly depends on the 4$a$ site irrespective of the structures.
In this discussion, therefore, in addition to the AF coupling between the Nd-Fe blocks, we paid attention to the magnetic interactions that arose from the 4$a$ Fe layers in Fig.~\ref{f2} because the magnetic moments of these Fe atoms reached $2.3\mu_{\mathrm{B}}/{\rm atom}$ (see Table~\ref{tbl:moment}) in both the FM and AF states. %, and the magnetic coupling should be considered. 
To estimate the magnetic coupling related to these 4$a$ Fe layers, we assumed the non-collinear magnetic structure exhibited in Fig.~\ref{f8} (a), where the magnetic moments of the 4$a$ Fe layers were rotated by an angle $\theta$ from those of Nd-Fe blocks. 
Figure~\ref{f8}~(b) shows the calculated energy of Nd$_{6}$Fe$_{14}$ as a function of $\theta$. 
The atomic position was fixed at the relaxed value of the FM state in Fig.~\ref{f2}, which corresponded to the $\theta = 0$ condition in Fig.~\ref{f8}~(a). 
We found that the calculated energy became unstable with increasing $\theta$; this result indicates that parallel magnetic couplings exist between the Nd-Fe blocks and the 4$a$ Fe layers. 
From the energy difference between $\theta = 0^{\circ}$ and $\theta = 180^{\circ} \sim 0.4\ \rm eV/unitcell$ in Fig.~\ref{f8}~(b),
the parallel magnetic coupling can be evaluated as $J_{\mathrm{NdFe/Fe}} \sim -12.5\ {\rm mJ/m^2}$ by using Eq.~\eqref{eq:2}.
Note that four interfaces between the Nd-Fe blocks and the Fe layers exist in the case of Fig.~\ref{f8} (a).
%
%\red{\sout{
%The FM state has parallel couplings between all neighboring Nd-Fe blocks and Fe layers, whereas the AF state has both parallel and anti-parallel couplings between the neighboring Nd-Fe blocks and Fe layers in a ratio of $1:1$ in the unit cell. 
%Therefore, from the energy difference between $\theta = 0^{\circ}$ and $\theta = 180^{\circ} \sim 0.4\ \rm eV/unitcell$ in Fig.~\ref{f8} (b), we demonstrated that part of $\Delta E = E_{\mathrm{FM}} - E_{\mathrm{AF}}$ originated from the magnetic coupling energy between the Nd-Fe blocks and Fe layers as $\Delta E \sim -0.2$ eV/unitcell. 
%This $\Delta E$ corresponds to $J_{\mathrm{NdFe/Fe}} \sim -0.05\ {\rm J/m^2}$.  
%}}
%
The parallel couplings compete with anti-parallel couplings between Nd-Fe blocks, $J_{\mathrm{NdFe/NdFe}}$, which have opposite sign and comparable magnitude (e.g., $J_{\mathrm{NdFe/NdFe}}\simeq J = \blue{8.7}\ {\rm mJ/m^2} $ for Nd$_6$Fe$_{13}$Ga in Table~\ref{t111}).
As a result, one can understand that the magnetic order of Nd$_6$Fe$_{14}$ is sensitive to the atomic positions (see Fig.~\ref{f6}~(c)).

From the above evaluation of the strength of magnetic couplings, 
we conclude that the cancellation of two kinds of magnetic coupling, $J_{\mathrm{NdFe/NdFe}}$ and $J_{\mathrm{NdFe/Fe}}$, induces the small energy difference observed between the FM and AF states in Nd$_{6}$Fe$_{14}$.
On the other hand, Nd$_{6}$Fe$_{13}$Ga has a single stable AF state because only the $J_{\mathrm{NdFe/NdFe}}$ has a considerable effect in the Nd$_{6}$Fe$_{13}$Ga structure.

In summary, we found that the substitution of Fe atoms in Nd$_6$Fe$_{14}$ by Ga atoms, leading to Nd$_6$Fe$_{13}$Ga, makes the AF state much stable, while Nd$_6$Fe$_{14}$ alloy favours ferromagnetic with a slight energy difference.  We also reveal that this stable AF state originates from the anti-parallel coupling between the neighbouring Nd-Fe blocks and non-magnetic doped Ga atoms contributes to the the stabilization of this magnetic state.  It may be possible to consider that the enhancement of coercivity of Nd-Fe-B magnets due to the addition of Ga atoms is related to the realization of AF state of Nd$_6$Fe$_{13}$Ga alloy. The formation of AF state in the GB phase could block the domain wall propagation or suppress the nucleation of reversed domains within the GB phases.

\section*{ACKNOWLEDGMENTS}
This work was supported by JST-CREST and the Elements Strategy Initiative Project (ESICMM) under the auspices of MEXT.  %(detail of foundation) 
Private communications with T. T. Sasaki, S. Doi, Y. Harashima, and S. Hirosawa are also appreciated.

\end{document}